\documentclass[conference]{IEEEtran}
\makeatletter
\def\ps@headings{%
\def\@oddhead{\mbox{}\scriptsize\rightmark \hfil \thepage}%
\def\@evenhead{\scriptsize\thepage \hfil \leftmark\mbox{}}%
\def\@oddfoot{}%
\def\@evenfoot{}}
\makeatother
\usepackage{cite}
\usepackage{amsmath,amssymb,amsfonts}
\usepackage{algorithmic}
\usepackage{graphicx}
\usepackage{textcomp}
\usepackage{bmpsize}
\usepackage{xcolor}
\usepackage{lipsum}
\usepackage{tabularx}
\usepackage{csquotes}

\newcolumntype{Y}{>{\centering\arraybackslash}X}
\usepackage{enumerate}
\usepackage{enumitem}
\usepackage[normalem]{ulem}

\MakeOuterQuote{"}
\def\BibTeX{{\rm B\kern-.05em{\sc i\kern-.025em b}\kern-.08em
    T\kern-.1667em\lower.7ex\hbox{E}\kern-.125emX}}
\begin{document}

\title{(Mis)perceptions and Engagement on Twitter: COVID-19 Vaccine Rumors on Efficacy and Mass Immunization Effort}

\author{\IEEEauthorblockN{Filipo Sharevski}
\IEEEauthorblockA{\textit{School of Computing} \\
\textit{DePaul University}\\
Chicago, IL \\
fsharevs@cdm.depaul.edu}
\and
\IEEEauthorblockN{Alice Huff}
\IEEEauthorblockA{\textit{School of Computing} \\
\textit{DePaul University}\\
Chicago, IL \\
arosent4@depaul.edu} \\
\and
\IEEEauthorblockN{Peter Jachim}
\IEEEauthorblockA{\textit{School of Computing} \\
\textit{DePaul University}\\
Chicago, IL \\
pjacim@depaul.edu} 
\and 
\IEEEauthorblockN{Emma Pieroni}
\IEEEauthorblockA{\textit{School of Computing} \\
\textit{DePaul University}\\
Chicago, IL \\
epironi@depaul.edu}}


\maketitle

\begin{abstract}
This paper reports the findings of a 606-participant study where we analyzed the perception and engagement effects of COVID-19 vaccine rumours on Twitter pertaining to (a) vaccine efficacy; and (b) mass immunization efforts in the United States. Misperceptions regarding vaccine efficacy were successfully induced through simple content alterations and the addition of popular anti COVID-19 hashtags to otherwise valid Twitter content. Twitter's misinformation contextual tags caused a ``backfire effect'' for the skeptic, vaccine-hesitant participants reinforcing their opposition stance. While the majority of the participants staunchly refrained from engaging with the COVID-19 rumours, the vaccine-hesitant ones were open to comment, re-tweet, like and share the vaccine efficacy rumors. We discuss the implications of our results in the context of broadening the effort for dispelling rumors about COVID-19 on social media.
\end{abstract}


\section{Introduction}
COVID-19, as an unprecedented threat to public health, has been surrounded with many unverified claims about: the virus propagation, mutations, long-term effects, vaccine development and mass immunization. These ambiguities allowed for misinformation and rumours to proliferate alongside public health authority's claims. Mindful of this ``infodemic,'' Twitter in time responded by issuing warnings - in form of contextual tags or interstitial covers - on Tweets deemed as spreading misinformation related to the COVID-19 pandemic and vaccine \cite{Roth, Kaiser}. However, there is no evidence that these warnings are as effective as anticipated. An early investigation of misinformation warnings on social media suggest that they may actually ``backfire,'' i.e. convince people to believe the misinformation even more than if the label were not there. One reason for this result is because the warnings were primarily focused on battling political misinformation, versus COVID-19, which has shown to be incredibly divisive \cite{Clayton}. This backfiring of warnings led users that believed whatever stance the misinformation declared to lose trust in the social media site's judgment. Instead of halting the spread, this lack of trust culminated in proliferating the misinformation. 

There are real world implications of (mis)information and unverified rumors having a direct impact on public health in terms of hesitancy to receive a COVID-19 vaccine. For this reason, we wanted to explore if (a) carefully altered Twitter content in the form of a rumour could cause misperceptions about the COVID-19 vaccination, and (b) initiate a desire to engage with this rumor, even in the presence of warnings in a form of contextual tags. We focused specifically on COVID-19 vaccines because of the relevance linked to development and deployment of several vaccines available at the time of the study in early 2021 \cite{shen21}. The other leading factor for testing vaccine specific content was the existing evidence of polarized discourse surrounding the federal vaccination effort on Twitter. Recent studies have shown that valid Twitter content on vaccines could be altered to cause a misperception about the relationship between vaccines and autism \cite{Sharevski0320}. Therefore, we sought to see how participants might respond to efficacy rumours and whether it would illicit a desire to engage in the discourse on Twitter.  Engagement, whether for purposes of negating the information or not, aids in further dissemination of that information, and early evidence showed that Tweets labeled as misinformation generate more engagement than regular Tweets \cite{Zannettou}. Because of the confines of Twitter interactions, we focused our study on comments, likes, re-tweets, and sharing actions. 

Our results suggest that people are overly sensitive to pessimistic rumours about the COVID-19 vaccine, as well as alternative hashtags \cite{Jachim}. Our results showed that it was sufficient for a Tweet to cause a misperception of otherwise valid content was not very accurate through the inclusion of popular alternative hashtags \#COVIDIOT and \#covidhoax. The participants in our study were also unable to shed their staunch notions about general vaccination efficacy when interpreting COVID-19 vaccination information on Twitter. The participants who have existing skepticism of the likelihood of a successful COVID-19 vaccine being produced (``vaccine hesitant participants''), were more inclined to accept a pessimistic alteration of COVID-19 vaccine content. The findings of our study add more evidence to the phenomenon of belief echoes on Twitter. We found that despite the contextual tags, there was sustained hesitancy to receive a vaccination personally or for children to receive one. Additionally, pre-existing beliefs about COVID-19 rumors, alternative narratives and misinformation were unaltered by the contextual tags. 



\section{COVID-19 Content on Twitter}
As one of the mainstream social media sites in the United States, Twitter became a battleground of the COVID-19 online discourse. This heated discourse required someone to assume the role of taming uncertainties. This duty fell on public healthcare authorities like the Centers for Disease Control (CDC) and the World Health Organization(WHO). Tasked with the righting of wrongs, public health officials timidly and cautiously joined the battle. But because of the initial lack of information surrounding COVID-19, they did so inconsistently, e.g. expressing reservations about effectiveness of masks to prevent the spread of the virus, later changing their view to proclaim masks' prevention efficacy \cite{ike}. Accordingly, this provided an opportunity to hijack the COVID-19 discourse.  It allowed dissenting users to spread rumors and disinformation regarding the virus, using the official's mistakes as fodder for their defense.  Due to the  majority of the public being at home with unlimited Internet access and time to kill, the discourse spread like wildfire ~\cite{Frenkel}. 

Twitter was initially hesitant to implement hard moderation (account bans) knowing dissenting users' valid argument for protection of free speech. Instead Twitter opted to suspend accounts on the grounds that content was violating the platform's terms of use. This in and of itself was an enormous task to undertake due to the amount of nuanced material to comb through. However, COVID-19 misinformation quickly became an ``infodemic,'' which forced the platform to monitor COVID-19 content for false or misleading information that was not corroborated by public health authorities or subject matter experts. Their attempt was to apply warnings in a form of interstitial covers or contextual tags underneath  unverified information \cite{Roth}. The supposed aim of these warnings is to reduce misleading or harmful information that could incite people to action and cause widespread panic, health anxiety, and fear that could lead to social unrest or large-scale disorder. 

However, one study found that Twitter's content with contextual tags generated more action than content without said labels \cite{Zannettou}. Meaning that the misinformation was spreading more due to the tag. Despite the public health risk, the study found a mere 1\% of the Tweets gathered (a total of 18,765 Tweets) were contextually tag with a COVID-19 warning. A number of these 187 some Tweets were found to be mislabeled simply because they contained the words ``oxygen'' and ``frequency.'' One such Tweet specifically was attempting to show the failures of the soft moderation for COVID-19 misinformation and invited others to test the keywords as well, i.e. by writing about mountain climbing ``oxygen'' levels and ``frequency'' to monitor gear. Another study found that a number of users did not trust the soft moderation intervention because it opposed their personal beliefs. Consequently, they felt that Twitter itself was biased and purposefully mislabeling valid content ~\cite{Geeng20}. 

The effort to tame the uncertainty surrounding COVID-19 and related vaccinations is a convoluted affair. Even with the attempt of soft moderation to emphasize invalid COVID-19 information, there exists the possibility for undetected circulation of COVID-19 misinformation or at least unverified rumors. These realizations led us to question the probability of bad actors responding to this demand for information through intentional spreading of rumors regarding COVID-19 on the social media platform Twitter. In order to evaluate the results of this threat we chose to analyze participants' reactions to altered content of Tweets as well as implementation of the contextual tags to rumors. We were also interested in investigating the level of engagement of Twitter users initiated by the perception of COVID-19 vaccination information pertaining to (a) vaccine efficacy; and (b) mass immunization effort.

\section{Research Study}
\subsection{COVID-19 Vaccine Misperceptions}
We set to examine the possibility for inducing misperceptions regarding the efficacy of the COVID-19 vaccines as well as the political context of the COVID-19 mass immunization. We selected two verified content Tweets to act as the controls (Figure 1 and Figure 3). The first Tweet seen in Figure 1 was a Tweet reporting the efficacy of the Oxford/Astra-Zeneca COVID-19 vaccines. This content was selected owing to the controversy surrounding the large-scale trials for this particular COVID-19 vaccine, its diminished effectiveness against new variants, as well as the mixed interpretation of the results for elderly \cite{wordsworth21}. By the time of the study, this vaccine has not received an approval by the Food and Drug Administration \cite{burgos}. This controversy created a polarized debate on Twitter and we explored if an alteration of the Tweet feeding into the downplay of its' effectiveness, shown in Figure 2, would suffice in affecting the perceived accuracy of the content. We tested the following hypothesis, accordingly: 

\vspace{0.5em}

\begin{itemize}[leftmargin=*, label={}]
\itemsep 1em
    \item $H1$: There will be no difference in the perceived accuracy between an altered Tweet containing \textit{misleading} information about the effectiveness of a COVID-19 vaccine relative to a Tweet containing \textit{valid} information about the effectiveness of a COVID-19 vaccine. 
\end{itemize}

To remove any bias or control for the ``influencer'' effect, all Tweets tested appear to come from a verified account named ``VaccinateNow'' and indicate a relatively high level of interaction with 15.3k re-tweets, 17.2 quotations, and 6.8K likes. This level of engagement is appropriate when compared to comparable Tweets with important COVID-19 vaccine information previously observed on Twitter by \cite{Zannettou}. For the opposing Tweet, the software from \cite{Sharevski0320} was utilized to swap the word ``robust'' with the word ``mild,'' to correspond to the differences in responses with the administration of full and half doses \cite{Callaway}. The software also negated the word ``could'' to ``couldn't'' and inserted the word ``lasting'' before the word ``immunity'' to emphasize the lack of evidence about the length of the immunity provided by this particular COVID-19 vaccine at the time of the study \cite{AnsweringPatientsQuestions}. The software also inserted two trending hashtags among top alternative COVID-19 Twitter users, \#COVIDIOT and \#covidhoax  \cite{Jachim, Chen}.  

\begin{figure}[htbp]
  \centering
  \includegraphics[width=0.8\linewidth]{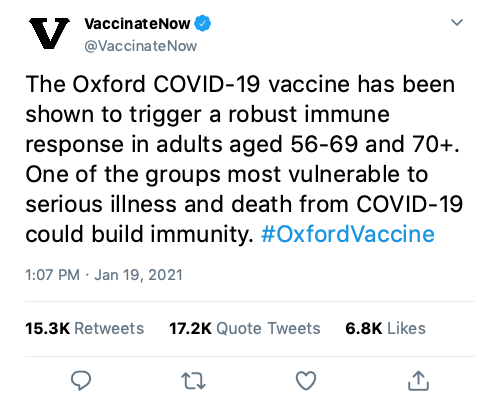}
  \caption{Verified Vaccine Efficacy Information Tweet.}
\end{figure}

\begin{figure}[htbp]
  \centering
  \includegraphics[width=0.8\linewidth]{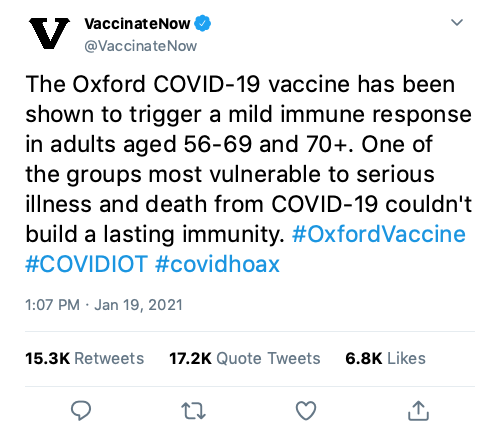}
  \caption{Altered Efficacy Information Tweet.}
\end{figure}

The misinformation labels on Twitter gained widespread attention with the soft (and later hard) moderation of political content \cite{Roth}. Twitter applied a similar approach of soft moderation to any unverified claim about the COVID-19 vaccines by applying labels with an exclamation mark and a link where users can ``get the facts about COVID-19.'' With the stark political division over the federal COVID-19 mass immunization \cite{O'Keefe}, we wanted to test the effect of label alteration in addition to the altering of Twitter COVID-19 vaccine content. We selected a Tweet, shown in Figure 3, reporting the intentions of the president-elect Joe Biden to drop the name ``Operation Warp Speed'' from the federal vaccine effort to ``combat the populist management of the COVID-19 pandemic by the previous administration of Donald Trump'' \cite{mckee} (we are apolitical as researchers and take no preference in political figures). Naturally, this turned into ammunition for sustaining the political/mass vaccination on Twitter. We explored if an alternation of the Tweet -- dropping the key word ``name,'' shown in Figure 4 -- might cause confusion that the effort for mass vaccination under the new administration is in jeopardy. We also explored a variation of the modified misinformation Tweet with the addition of a contextual tag (Figure 5) in order to see if users will heed a misinformation warning. We tested the following hypotheses, accordingly:  

\vspace{0.5em}

\begin{itemize}[leftmargin=*, label={}]
\itemsep 1em
    \item $H2$: There will be no difference in the perceived accuracy between an altered Tweet containing \textit{misleading} information about the COVID-19 mass immunization relative to a Tweet containing \textit{valid} information about the COVID-19 mass immunization. 
    
    \item $H3$: There will be no difference in the perceived accuracy between an altered Tweet containing \textit{misleading} information and a COVID-19 misinformation contextual tag relative to a Tweet containing \textit{valid} information about the COVID-19 mass immunization. 
    
\end{itemize}

\begin{figure}[htbp]
  \centering
  \includegraphics[width=0.8\linewidth]{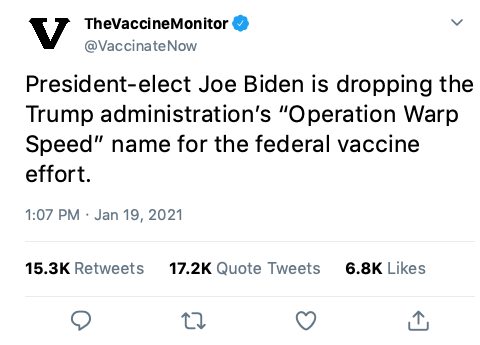}
  \caption{Verified Mass Immunization Information Tweet.}
\end{figure}

\begin{figure}[htbp]
  \centering
  \includegraphics[width=0.8\linewidth]{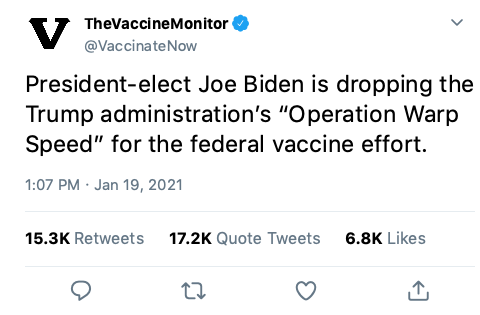}
  \caption{Altered Mass Immunization Information Tweet \textit{Without} a Contextual Tag}
\end{figure}

\begin{figure}[htbp]
  \centering
  \includegraphics[width=0.9\linewidth]{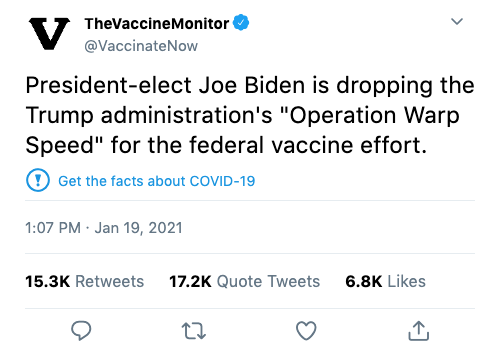}
  \caption{Altered Mass Immunization Information Tweet \textit{With} a Contextual Tag}
\end{figure}

Because the Tweet's content is on COVID-19 vaccination, we tested the relationship between one's hesitancy to receive a COVID-19 vaccination (personally and a vaccination for children) as well as their beliefs on production of safe and effective vaccines and the perceived accuracy of the Tweets in Figure 1 - 5 with the following hypotheses:

\vspace{0.5em}

\begin{itemize}[leftmargin=*, label={}]
\itemsep 1em
    \item $H4_{1}$: There will be no difference in the perceived accuracy of an altered Tweet containing \textit{misleading} information about the COVID-19 vaccines between Twitter users that are personally hesitant and users that are willing to receive the COVID-19 vaccine for themselves.
    
    \item $H4_{1}$: There will be no difference in the perceived accuracy of an altered Tweet containing \textit{misleading} information about the COVID-19 vaccines between Twitter users that are hesitant and users that are willing to administer the COVID-19 vaccine to children.
    
    \item $H4_{3}$: There will be no difference in the perceived accuracy of a Tweet containing \textit{valid} information about the COVID-19 vaccines between Twitter users that believe a safe and effective COVID-19 vaccine is possible and the users that believe that's not possible. 
\end{itemize}



    
 

\subsection{COVID-19 Vaccine Twitter Engagement}
Engagement with soft-moderated Twitter content and misinformation content was found to be high among Twitter users \cite{Zannettou}. Therefore, we also explored the intended engagement with the Tweets in Figure 1 - 5. We assessed the likelihood of commenting, re-tweeting, liking, and sharing the Tweets to test the following hypotheses: 

\vspace{0.5em}

\begin{itemize}[leftmargin=*, label={}]
\itemsep 1em
    \item $H5$: There will be no difference in the likelihood for engagement (commenting, re-tweeting, liking, and sharing) between an altered Tweet containing \textit{misleading} information about the COVID-19 vaccines relative to a Tweet containing \textit{valid} information about the COVID-19  vaccines. 

    \item $H6_{1}$: There will be no difference in the engagement (commenting, re-tweeting, liking, and sharing) with an altered Tweet containing \textit{misleading} information about the COVID-19 vaccines between hesitant and non-hesitant Twitter users, both personally and for children. 
    
    \item $H6_{2}$: There will be no difference in the engagement (commenting, re-tweeting, liking, and sharing) with a Tweet containing \textit{valid} information about the COVID-19 vaccines between hesitant and non-hesitant Twitter users, both personally and for children. 
    
    \item $H7_{1}$: There will be no difference in the engagement (commenting, re-tweeting, liking, and sharing) with an altered Tweet containing \textit{misleading} information about the COVID-19 vaccines between Twitter users that believe a safe and effective COVID-19 vaccine is possible and the users that believe that's not possible.
    
    \item $H7_{2}$: There will be no difference in the engagement (commenting, re-tweeting, liking, and sharing) with a Tweet containing \textit{valid} information about the COVID-19 vaccines between Twitter users that believe a safe and effective COVID-19 vaccine is possible and the users that believe that's not possible. 
    
\end{itemize}

\subsection{Sampling and Instrumentation}
Prior to initiating the study, we received approval from our local Institutional Review Board. We set to sample a population that met the following base requirements: participant was 18 years old or above, was a Twitter user, and has encountered at least one Tweet in their Twitter feed that relates to COVID-19 vaccines.  These requirements were implemented using metric tools as part of survey posting on Prolific and "Human Intelligence Tasks" (HITs) posting on Amazon Mechanical Turk (``MTurk''). We crafted the content of the Tweets to be relevant to the participants, such that they may wish to meaningfully engage with the Tweet’s content (i.e., their responses are not arbitrary). 

To assess the perceived accuracy we used the questionnaire from \cite{Clayton} for each of the Tweets on a 4-point Likert scale (1-not at all accurate, 2-not very accurate, 3-somewhat accurate, 4-very accurate). To assess participants’ hesitancy and beliefs regarding the COVID-19 vaccine, we used the questionnaire from \cite{Biasio}. To assess the subjective attitudes we asked if the participants (a) expect efficacious vaccine to be developed (Yes/No); (b) will receive a COVID-19 vaccine (Yes/No/I Don't Know); and (c) if children should receive a COVID-19 vaccine too (Yes/No). To gauge whether participants’ would engage with the Tweet, we used a standardized questionnaire for Twitter engagement on a 7-point Likert scale (1-extremely likely; 7-extremely unlikely) \cite{Sharevski0320}. We utilized an experimental design where participants were randomized into one of five groups: (1) verified vaccine efficacy information Tweet; (2) altered vaccine efficacy information Tweet; (3) verified mass immunization information Tweet (4) altered mass immunization information Tweet \textit{without} a warning tag; (5) altered mass immunization information Tweet \textit{with} a warning tag.

\section{Results}
We conducted an online survey (N = 606) in January and February 2021. The break down of participants' sex were as follows: 54\% male, 43.9\% female, and 2.1\% participants identified as non-cis, non-binary or preferring not to answer. The age brackets were distributed as follows: 20.0\% [18 - 24], 37.5\% [25 - 34], 25.5\% [35 - 44] and 16.8\% [45 - above]. The political leaning of the sample was skewed towards liberals: 51.8\% participants identified as liberal-leaning, 22.4\% identified as moderate and 25.8\% participants identified as conservative-leaning. 

\subsection{COVID-19 Vaccine Perceptions on Efficacy}
Initially we hypothesized that there would be no difference in the perceived accuracy between an altered Tweet containing \textit{misleading} information and an original Tweet containing \textit{valid} information about the effectiveness of a COVID-19 vaccine. We found a significant difference in the perceived accuracy between the Tweets in Figure 1 and Figure 2 ($U = 981$, $p = .000^{*}, (\alpha = 0.05)$, Cohen's $d =0.832$ large). Based on this result, we reject our first hypothesis and accept the alternative where the contextual rewording was perceived as ``not at all accurate,'' whereas the original Tweet was perceived as ``somewhat accurate'' on average. Perception of accuracy was altered through 1) swapping the word ``robust'' with ``mild,'' 2) the rewording to emphasize the lack of evidence of lasting immunization, 3) implementation of the most popular COVID-19 alternative hashtags (\#COVIDIOT and \#covidhoax). Either the participants in the altered Tweet group were overly sensitive to a pessimistic COVID-19 vaccine outlook, or a simple inclusion of alternative hashtags signaled ``opposition, fake news'' (recalling our liberal-leaning sample) \cite{Pennycook}. 

Hesitancy to receive the vaccine proved to be a decisive factor in how the altered Tweet was perceived. We found a statistically significant difference between the pro-vaccination and anti-vaccination participants for both condition of receiving a COVID-19 vaccination personally and administering one to children ($U = 567$, $p = .017^{*}, (\alpha = 0.05)$, Cohen's $d =0.4$ medium) and  ($U = 954$, $p = .033^{*}, (\alpha = 0.05)$, Cohen's $d =0.4$ medium), respectively. Rejecting $H4_{1}$ and $H4_{2}$ hypotheses, we accept the alternative hypothesis that one's hesitancy to receive a vaccine factors into how COVID-19 information is perceived. The hesitant participants perceived the altered Tweet as ``somewhat accurate,'' while the pro-vaccination participants perceived it as ``not very accurate.'' This comes as no surprise given that the pessimistic rewording confirms the suspicions of the overall COVID-19 vaccination effort \cite{kreps}. To underline this point, we also had to reject the $H4_{3}$ hypothesis and accept the alternative. The alternative suggested that the participants with a pre-existing skepticism of the possibility of a safe and efficacious vaccine also deemed the alternative Tweet as ''somewhat accurate'' ($U = 266$, $p = .030^{*}, (\alpha = 0.05)$, Cohen's $d =0.3$ small). 

\subsection{COVID-19 Vaccine Perceptions on Mass Immunization}
To investigate the possibility for misperceptions further we next hypothesized that there will be no difference in the perceived accuracy between an altered and original Tweet on the topic of mass immunization. For the second test of misperception of COVID-19 vaccines we opted to test a more politicized Tweet. COVID-19 was one of the main focal points of the political battle during and after the U.S. elections in 2020 that naturally flooded over to Twitter \cite{Jachim}. Therefore, we tested a Tweet regarding the new administration's intentions for renaming ``Operation Warp Speed,'' the Department of Defense's effort for rapid U.S. mass immunization \cite{dod}. Here we took a slightly more adversarial approach in attempting to muddy the waters about what the President-elect had reported to drop - just the name or perhaps the entire operation, given his open criticism of the operation overall \cite{Kaplan}. 

We found a significant difference in the perceived accuracy between the Tweets in Figure 3 and Figure 4 ($U = 1845.5$, $p = .023^{*}, (\alpha = 0.05)$, , Cohen's $d =0.619$ medium). Based on this result, we reject our second hypothesis and accept the alternative one where the contextual rewording was perceived as ``not very accurate,'' whereas the original Tweet was on average considered ``somewhat accurate.'' This is a promising result suggesting that Twitter users in our liberal-leaning sample can accurately assess an attempt for spreading rumours about this vital operation for mass immunization. Perhaps this is an expected result given that liberal-leaning, and possibly moderate users, are sensitive to any attempt to tarnish the actions of Donald Trump, who is widely accepted as the top misinformation machine over the last four years\cite{Jachim}. Or these participants closely monitor mainstream media compared to their conservative counterparts \cite{Ferrara}. 

Indeed, the participants heeded the contextual tag applied to the altered variant of the Tweet (Figure 5). We found a significant result in the perceived accuracy for the labeled Tweet and the original Tweet in Figure 3 ($U = 2825$, $p = .002^{*}, (\alpha = 0.05)$, Cohen's $d =0.532$ medium). We rejected the third hypothesis and accepted the alternative, that the warning tag indeed nudged the participants to perceive the Tweet as ``not at all accurate.'' This evidence goes along with the observation that misinformation labels on social media works, if that label aligns with one's biases and receptivity to the content at stake \cite{Clayton}. This finding indicates that the liberal-leaning and moderate participants trust Twitter and the soft moderation of COVID-19 vaccination content. This is contrary to the evidence of opposition sentiment, that did not trust the soft moderation intervention and felt that Twitter itself was biased and mislabeling content \cite{Geeng}. 
As we previously found, hesitancy to receive the vaccine again proved to be a decisive factor in how the misinformation labeled Tweet was perceived. We found a statistically significant difference between the pro-vaccination and anti-vaccination participants for both condition of receiving a COVID-19 vaccination personally and administering one to children ($U = 453$, $p = .033^{*}, (\alpha = 0.05)$, Cohen's $d =0.4$ medium) and  ($U = 608$, $p = .014^{*}, (\alpha = 0.05)$, Cohen's $d =0.233$ small), respectively. Rejecting $H4_{1}$ and $H4_{2}$ hypotheses, we accept the alternative hypothesis that one's hesitancy factors into how COVID-19 information is perceived. The vaccine hesitant participants perceived the altered Tweet as ``somewhat accurate,'' while the pro-vaccination participants viewed it as ``not very accurate.'' Again, this breakdown reveals that heeding a misinformation warning relies on the biases regarding the content of the Tweet \cite{Clayton}. We also had to reject the $H4_{3}$ hypothesis and accept the alternative one suggesting that the vaccine hesitant participants deemed the altered Tweet claiming Operation Warp Speed as ``somewhat accurate'' despite the soft moderation warning ($U = 266$, $p = .030^{*}, (\alpha = 0.05)$, Cohen's $d =0.3$ small).

\subsection{COVID-19 Vaccine Twitter Engagement}
To test the likelihood of engagement with each of our Tweets in the study, we hypothesised that there will be no difference in level of commenting, re-tweeting, liking, and sharing between an altered and the original versions of the Tweets in Figures 1-5. Comparing the engagement with the Tweets on COVID-19 vaccine efficacy (Figure 1 and Figure 2), we observed a statistical difference in the case of re-tweeting ($U = 986.1$, $p = .002^{*}, (\alpha = 0.05)$, Cohen's $d =0.2$ small), liking ($U = 165.9$, $p = .000^{*}, (\alpha = 0.05)$, Cohen's $d =0.23$ small), and sharing ($U = 1007$, $p = .002^{*}, (\alpha = 0.05)$, Cohen's $d =0.267$ small0), where the altered Tweet was ``extremely unlikely'' to be engaged with, compared with the ``somewhat unlikely'' with the original Tweet. Comparing the engagement with the Tweets on the COVID-19 mass immunization, we didn't observe any statistical difference. 

In contrast to the evidence of high engagement with alternative and soft moderated Tweets \cite{Zannettou}, our sample appeared quite reserved in terms of engagement with the content offered. The unwillingness to engage with the twitter rumors is otherwise consistent with the spiral-of-silence effect observed for the general vaccination debate in \cite{Sharevski0320}. The evidence of high engagement was reported in the context of mocking the original poster and attempting to correct or debunk the perceived misinformation. However, our sample group was observed to have no intention of commenting or replying to either of the altered Tweets directly in order to take said actions. 

This could be a result of social network fatigue being a year into social media coverage of COVID-19 \cite{LIU21}. Otherwise, we observed a significant difference in engagement when we controlled for the hesitancy of COVID-19 vaccination, both personally and for children. The vaccine hesitant participants were ``somewhat likely'' to comment, re-tweet, like or share the altered Tweet seen in Figure 2. The ones with little belief for a production of safe and efficacious vaccines were also significantly more inclined to comment and re-tweet the altered Figure 2 Tweet, but not to like or share it. Rejecting the $H6_{1}$ and $H7_{1}$ hypotheses only for the pessimistic case, but not the other alterations including the soft moderated Tweet, we suspect is due to subjective interpretation of the content, as we noted previously.  

\section{Discussion}
\subsection{Broader Context of the Results}
In this study, we attempted to manufacture ``misinformation'' that essentially categorizes as a rumor more so than any of the other alternative narrative types \cite{Caulfield}. The deliberate choice for a nuanced modification of small, seemingly inconsequential changes in the content was made to capture the zeitgeist of uncertainty surrounding COVID-19 vaccination. This is especially prevalent in the politicization of the mass immunization effort. In order to capture the perceptions and the intent for engagement with content that is not clear cut, we chose this more nuanced approach versus blatant misinformation like the predominant COVID-19 vaccine sentiment on Parler \cite{Pieroni}. Yet another study of testing the claim that ``the COVID-19 vaccine will infect you with HIV'' with our liberal-leaning, dominantly young sample, would not have adequately yielded the perception whims and engagement avoidance proclivity. Finally, the more divisive misinformation might not have accurately assessed the vaccine hesitant participants' true inclinations and ways of interpreting information that fits broadly into a skeptic outlook of the mass COVID-19 vaccination.  

In terms of perceptions of COVID-19 rumors as Twitter content, this study helped conclude that existing biases, such as reservations of government's intention or skepticism of vaccine efficacy, have an impact on perception. Those with pre-existing skepticism and a hesitancy to personally receive a COVID-19 vaccine or administer one to children were more accepting of the altered Tweets presented. Those with no hesitancy in receiving a COVID-19 vaccine, and who believed in efficacy of existing vaccines, in contrast decisively did not accept the rumors. This example shows the effect of rumor propagation via echo chambers on social media \cite{choi20}. 

While other studies implied that there would be heavy engagement with misinformation, even for those who may disagree or not believe the misinformation, we found that most Twitter users in our sample were unlikely to comment, like, re-weet or share altered Tweets. Perhaps the rumors give people a pause because they cannot immediately infer the weaponizing value of the Tweet for their expression on Twitter versus the clear cut misinformation like ``5G causes coronavirus.'' The study showed that only those with skepticism were willing to engage with the Tweets. Another reason why the majority of the sample group, beyond the spiral-of-silence, may have been less inclined to engage may have to do with ``social network fatigue'' \cite{LIU21}.  The outcome of this fatigue is that social network users may skim or skip irrelevant information or even avoid some information, and exhibit ignoring and avoidance behaviors.

\subsection{Usable Security Implications}
We also focused on soft moderation, as an early effort to regulate the COVID-19 information, since misinformation could have ramifications beyond the microblogging sphere for the health of the general public. The majority of our participants were receptive to the soft moderation, which is a promising result and we acknowledge and support this effort for misinformation warnings. That being said, young liberal-leaning people do not make up the whole of the population. The concern we have is with the minority of our sample that chose to ignore these warnings. Efforts have been invested in increasing the clarity of the messages and design of soft moderation warnings to attract attention and motivate users. However, old habits die hard and habituation is a complex problem transcending security designs. 

The contextual tag implemented by Twitter in blue font appears as a banner after the Tweet content and any images/links with a favicon of an encircled exclamation point stating ``Get the facts about COVID-19,'' which redirects users to verified public health official's information. This design and formatting can be observed as innocuous, and does not explicitly address that the Tweet's content aims to mislead users about COVID-19 or its vaccines. A similar visual formatting is used for labeling Tweets with unverified political claims, e.g. ``Get the facts about mail-in ballots'' \cite{Roth}. Research has shown that even if people are exposed to misinformation multiple times, it can alter their memories \cite{nahleen}. For this reason, it may be worth exploring the potential benefits of adding the contextual tag above the content versus below it to assess if it hinders users from reading the misinformation.

Twitter just recently decided to up the ante in tagging intentional content-level exploits about COVID-19. The moderation is changing to a hybrid between hard and soft moderation, with a ``striking system'' that results in an ultimate ban from the platform after 5 strikes \cite{TwitterSafety}. It is interesting to research both the positive and negative externalities of this hybrid moderation effort. The hybrid moderation might restore the balance on Twitter, but further push the polarization between platforms that was already observed with the formation of a sizable Parler community of skeptic, COVID-19 vaccine-hesitant communities on Parler ~\cite{Pieroni}.

\subsection{Future Research}
Extensive further research should be done investigating the full ramifications of misinformation and soft/hybrid moderation by social media platforms. A promising line of research is the combination of soft and hard moderation, given that Twitter has exercised the right to ban or suspend accounts indefinitely that have been tagged for misinformation in the past, like in the case of Donald Trump. Twitter is going to implement a strike system for misinformation Tweets \cite{TwitterSafety} and research could probe the warning tagging/covering algorithm and reverse engineer it to find if a strike system will be more effective in curbing users posting misinformation, before the account gets permanently banned. 

Soft moderated content is typically closely related to trolling content, so there is room for exploration of this relationship, such as understanding if warning tagged/covered Tweets provoke emotional response and if so, what kind. Similar to research conducted on the evolution of COVID-19 information, the warning tagging/covering could be associated with an evolution of political information operations on Twitter. It would be beneficial to trace the relationship between actual users and social bots rigging the engagement metrics as in the previous vaccine debates on Twitter \cite{Sharevski0320}. 

\subsection{Scope Limitations}
We used Tweets that were tied to a particular vaccine vendor and a single decision regarding the public relations of United States mass immunization efforts during the period of January-February 2021. Twitter content tested did not include the actual operational changes promised or undertaken by then President Biden, which could be perceived with a different level of accuracy after a certain period of time. It is possible that other vaccines from various non-US vendors like Sanofi, Sinopharm or Galeneya, could yield different perception of accuracy or strength of soft moderation. Overall, the findings in the present study may be specific to the alterations we tested, and cannot be generalized to other alterations, for example swapping the word ``Warp'' with `Top'' in the second Tweet. Participants who are frequent social media users in general may be desensitized to the information presented in the Tweets. Which seems likely considering the breakdown of political leanings and age bracket of the majority of the test sample. The participants may also have been biased from heightened exposure to mainstream media and social network information about COVID-19 vaccines and the Biden administration mass immunization efforts. Both of these factors may have limited participants' perceptions and desire to engage with the content presented irrespective of the alterations. 


\section {Conclusion}
\label{sec:8}
This study showed that a majority of participants were able to recognize rumors and had no inclination to engage with those rumors (they did not have a desire to like, comment, re-tweet, or share the content). However, the majority of the participants identified as pro-vaccine and our sample was skewed towards the liberal-leaning.  These existing biases likely impacted their perceptions and lack of motivation to engage. The more participants who showed hesitancy to personally receiving a vaccine or administering them to children found the rumors more ``accurate'' and had more of an appetite to engage with the altered Tweets. While the skeptic, vaccine-hesitant participants were in the minority in our study and might well be a minority on twitter, that is not necessarily the makeup of the entire social media user base, especially considering alternative platforms like Parler. It is important to consider the potential consequences for overall public health of the soft moderation. While it is reassuring that majority of the participants had enough COVID-19 literacy to recognize rumors, this does not solve the problem of misinformation spreading on Twitter and the spill over on other alternative platforms. Nor does it alter the existing opinions of the opposing population with hesitancy to receive vaccination. Therefore, we an improved soft/hybrid moderation should be extensively explored in future towards effective eradication of valid COVID-19 vaccine information.

\bibliographystyle{IEEEtran}
\bibliography{sample-base}

\begin{thebibliography}{10}
\providecommand{\url}[1]{#1}
\csname url@samestyle\endcsname
\providecommand{\newblock}{\relax}
\providecommand{\bibinfo}[2]{#2}
\providecommand{\BIBentrySTDinterwordspacing}{\spaceskip=0pt\relax}
\providecommand{\BIBentryALTinterwordstretchfactor}{4}
\providecommand{\BIBentryALTinterwordspacing}{\spaceskip=\fontdimen2\font plus
\BIBentryALTinterwordstretchfactor\fontdimen3\font minus
  \fontdimen4\font\relax}
\providecommand{\BIBforeignlanguage}[2]{{%
\expandafter\ifx\csname l@#1\endcsname\relax
\typeout{** WARNING: IEEEtran.bst: No hyphenation pattern has been}%
\typeout{** loaded for the language `#1'. Using the pattern for}%
\typeout{** the default language instead.}%
\else
\language=\csname l@#1\endcsname
\fi
#2}}
\providecommand{\BIBdecl}{\relax}
\BIBdecl

\bibitem{Roth}
\BIBentryALTinterwordspacing
Y.~Roth and N.~Pickles, ``Updating our approach to misleading information,''
  2020. [Online]. Available:
  \url{https://blog.twitter.com/en\_us/topics/product/2020/updating-our-approach-to-misleading-information.html}
\BIBentrySTDinterwordspacing

\bibitem{Kaiser}
B.~Kaiser, J.~Wei, E.~Lucherini, K.~Lee, J.~N. Matias, and J.~Mayer, ``Adapting
  security warnings to counter online disinformation,'' 2020,
  \url{https://arxiv.org/abs/2008.10772}.

\bibitem{Clayton}
K.~Clayton, S.~Blair, J.~A. Busam, S.~Forstner, J.~Glance, G.~Green, A.~Kawata,
  A.~Kovvuri, J.~Martin, E.~Morgan \emph{et~al.}, ``Real solutions for fake
  news? measuring the effectiveness of general warnings and fact-check tags in
  reducing belief in false stories on social media,'' \emph{Political
  Behavior}, pp. 1--23, 2019.

\bibitem{shen21}
A.~K. Shen, R.~Hughes~Iv, E.~DeWald, S.~Rosenbaum, A.~Pisani, and W.~Orenstein,
  ``Ensuring equitable access to covid-19 vaccines in the us: Current system
  challenges and opportunities: Analysis examines ensuring equitable access to
  covid-19 vaccines.'' \emph{Health Affairs}, pp. 10--1377, 2021.

\bibitem{Sharevski0320}
\BIBentryALTinterwordspacing
F.~Sharevski, P.~Jachim, and K.~Florek, ``Tweet or not to tweet: Covertly
  manipulating a twitter debate on vaccines using malware-induced
  misperceptions,'' \emph{arXiv}, vol. 2003.12093v1, March 2020. [Online].
  Available: \url{https://arxiv.org/pdf/2003.12093.pdf}
\BIBentrySTDinterwordspacing

\bibitem{Zannettou}
\BIBentryALTinterwordspacing
S.~Zannettou, ``"i won the election!":an empirical analysis of soft moderation
  interventions on twitter,'' \emph{arXiv}, vol. 2101.07183v1, January 2021.
  [Online]. Available: \url{https://arxiv.org/pdf/2101.07183.pdf}
\BIBentrySTDinterwordspacing

\bibitem{Jachim}
\BIBentryALTinterwordspacing
P.~Jachim, F.~Sharevski, and P.~Treebridge, ``Trollhunter [evader]: Automated
  detection [evasion] of twitter trolls during the covid-19 pandemic,'' in
  \emph{New Security Paradigms Workshop 2020}, ser. NSPW '20.\hskip 1em plus
  0.5em minus 0.4em\relax New York, NY, USA: Association for Computing
  Machinery, 2020, p. 59–75. [Online]. Available:
  \url{https://doi.org/10.1145/3442167.3442169}
\BIBentrySTDinterwordspacing

\bibitem{ike}
J.~D. Ike, H.~Bayerle, R.~A. Logan, and R.~M. Parker, ``Face masks: Their
  history and the values they communicate,'' \emph{Journal of Health
  Communication}, pp. 1--6, 2021.

\bibitem{Frenkel}
S.~Frenkel, M.~Abi-Habib, and J.~E. Barnes, ``{Russian Campaign Promotes
  Homegrown Vaccine and Undercuts Rivals},'' 2021,
  \url{https://www.nytimes.com/2021/02/05/technology/russia-covid-vaccine-disinformation.html}.

\bibitem{Geeng20}
\BIBentryALTinterwordspacing
C.~Geeng, T.~Francisco, J.~West, and F.~Roesner, ``Social media covid-19
  misinformation interventions viewed positively, but have limited impact,''
  \emph{arXiv}, vol. 2012.11055v1, December 2020. [Online]. Available:
  \url{https://arxiv.org/pdf/2012.11055.pdf}
\BIBentrySTDinterwordspacing

\bibitem{wordsworth21}
M.~Wordsworth, S.~Scott, S.~Gilbert, G.~Hunt, K.~Quinn, and C.~Vinuesa,
  ``Covid-19 vaccine:: Disappointing result: New data about the
  oxford-astrazeneca vaccine has cast doubt over how effective it might be
  against some forms of the covid-19 virus.'' 2021.

\bibitem{burgos}
R.~M. Burgos, M.~E. Badowski, E.~Drwiega, S.~Ghassemi, N.~Griffith, F.~Herald,
  M.~Johnson, R.~O. Smith, and S.~M. Michienzi, ``The race to a covid-19
  vaccine: opportunities and challenges in development and distribution,''
  \emph{Drugs in Context}, vol.~10, 2021.

\bibitem{Callaway}
{Callaway, Ewen}, ``{Why Oxford’s positive COVID vaccine results are puzzling
  scientists},'' November 2020,
  \url{https://www.nature.com/articles/d41586-020-03326-w}.

\bibitem{AnsweringPatientsQuestions}
{Centers for Disease Control and Prevention (CDC)}, ``{Answering Patients'
  Questions},'' January 2021,
  \url{https://www.cdc.gov/vaccines/covid-19/hcp/answering-questions.html}.

\bibitem{Chen}
E.~Chen, K.~Lerman, and E.~Ferrara, ``Tracking social media discourse about the
  covid-19 pandemic: Development of a public coronavirus twitter data set,''
  \emph{JMIR Public Health Surveillance}, vol.~6, no.~2, p. e19273, May 2020.

\bibitem{O'Keefe}
\BIBentryALTinterwordspacing
S.~O'Keefe, ``One in three americans would not get covid-19 vaccine,''
  \emph{Gallup}, 2020. [Online]. Available:
  \url{https://news.gallup.com/poll/317018/one-three-americans-not-covid-vaccine.aspx}
\BIBentrySTDinterwordspacing

\bibitem{mckee}
M.~McKee, A.~Gugushvili, J.~Koltai, and D.~Stuckler, ``Are populist leaders
  creating the conditions for the spread of covid-19?; comment on “a scoping
  review of populist radical right parties’ influence on welfare policy and
  its implications for population health in europe”,'' \emph{International
  journal of health policy and management}, 2020.

\bibitem{Biasio}
L.~R. Biasio, G.~Bonaccorsi, C.~Lorini, and S.~Pecorelli, ``Assessing covid-19
  vaccine literacy: a preliminary online survey,'' \emph{Human Vaccines \&
  Immunotherapeutics}, vol.~0, no.~0, pp. 1--9, 2020.

\bibitem{Pennycook}
G.~Pennycook and D.~G. Rand, ``Who falls for fake news? the roles of bullshit
  receptivity, overclaiming, familiarity, and analytic thinking,''
  \emph{Journal of Personality}, vol.~88, no.~2, pp. 185--200, 2020.

\bibitem{kreps}
S.~Kreps, S.~Prasad, J.~S. Brownstein, Y.~Hswen, B.~T. Garibaldi, B.~Zhang, and
  D.~L. Kriner, ``Factors associated with us adults’ likelihood of accepting
  covid-19 vaccination,'' \emph{JAMA network open}, vol.~3, no.~10, 2020.

\bibitem{dod}
\BIBentryALTinterwordspacing
{U.S. Department of Defense}, ``{Operation Warp Speed},'' 2021. [Online].
  Available:
  \url{https://www.defense.gov/Explore/Spotlight/Coronavirus/Operation-Warp-Speed/}
\BIBentrySTDinterwordspacing

\bibitem{Kaplan}
\BIBentryALTinterwordspacing
{Kaplan, Thomas and Robbins, Rebecca}, ``{Biden Criticizes Trump on Vaccine
  Distribution and Pledges to Pick Up Pace},'' 2020. [Online]. Available:
  \url{https://www.nytimes.com/2020/12/29/us/politics/biden-coronavirus-vaccines.html}
\BIBentrySTDinterwordspacing

\bibitem{Ferrara}
\BIBentryALTinterwordspacing
E.~Ferrara, H.~Chang, E.~Chen, G.~Muric, and J.~Patel, ``Characterizing social
  media manipulation in the 2020 u.s. presidential election,'' \emph{First
  Monday}, vol.~25, no.~11, pp. 1--13, Oct. 2020. [Online]. Available:
  \url{https://journals.uic.edu/ojs/index.php/fm/article/view/11431}
\BIBentrySTDinterwordspacing

\bibitem{Geeng}
C.~Geeng, T.~Francisco, J.~West, and F.~Roesner, ``Social media covid-19
  misinformation interventions viewed positively, but have limited impact,''
  2020.

\bibitem{LIU21}
\BIBentryALTinterwordspacing
H.~Liu, W.~Liu, V.~Yoganathan, and V.-S. Osburg, ``Covid-19 information
  overload and generation z's social media discontinuance intention during the
  pandemic lockdown,'' \emph{Technological Forecasting and Social Change}, vol.
  166, p. 120600, 2021. [Online]. Available:
  \url{https://www.sciencedirect.com/science/article/pii/S0040162521000329}
\BIBentrySTDinterwordspacing

\bibitem{Caulfield}
\BIBentryALTinterwordspacing
S.~Zannettou, T.~Caulfield, E.~De~Cristofaro, N.~Kourtelris, I.~Leontiadis,
  M.~Sirivianos, G.~Stringhini, and J.~Blackburn, ``The web centipede:
  Understanding how web communities influence each other through the lens of
  mainstream and alternative news sources,'' in \emph{Proceedings of the 2017
  Internet Measurement Conference}, ser. IMC '17.\hskip 1em plus 0.5em minus
  0.4em\relax New York, NY, USA: Association for Computing Machinery, 2017, p.
  405–417. [Online]. Available: \url{https://doi.org/10.1145/3131365.3131390}
\BIBentrySTDinterwordspacing

\bibitem{Pieroni}
E.~Peironi, P.~Jachim, N.~Jachim, and F.~Sharevski, ``Parlermonium: A
  data-driven ux design evaluation of the parler platform,'' in \emph{Critical
  Thinking in the Age of Misinformation CHI 2021}, 2021.

\bibitem{choi20}
D.~Choi, S.~Chun, H.~Oh, J.~Han \emph{et~al.}, ``Rumor propagation is amplified
  by echo chambers in social media,'' \emph{Scientific reports}, vol.~10,
  no.~1, pp. 1--10, 2020.

\bibitem{nahleen}
S.~Nahleen, D.~Strange, and M.~K. Takarangi, ``Does emotional or repeated
  misinformation increase memory distortion for a trauma analogue event?''
  \emph{Psychological Research}, pp. 1--13, 2020.

\bibitem{TwitterSafety}
{Twitter Safety}, ``Updates to our work on covid-19 vaccine misinformation.''

\end{thebibliography}

\end{document}